\documentclass[review]{elsarticle}

\usepackage{lineno,hyperref}
\modulolinenumbers[5]

\journal{Journal of \LaTeX\ Templates}

\def\NP{Nucl. Phys. }
\def\PR{Phys. Rev. }
\def\PRL{Phys. Rev. Lett. }
\def\PL{Phys. Lett. }

\def\epj{Eur. Phys. J. }

\begin{document}

\begin{frontmatter}

\title{Transverse wobbling motion in $^{134}$Ce and $^{136}$Nd}

\author{C. M. Petrache}
\author{S. Guo\fnref{guo}}
\address{Centre de Sciences Nucl\'eaires et Sciences de la Mati\`ere, CNRS/IN2P3, Universit\'{e} Paris-Saclay, B\^at. 104-108, 91405  Orsay, France}

\fntext[guo]{On leave from: Institute of Modern Physics, Chinese Academy of Sciences, Lanzhou, China} 

\cortext[mycorrespondingauthor]{Corresponding author}
\ead{petrache@csnsm.in2p3.fr}


\begin{abstract}
 The existence of one-phonon and possible two-phonon transverse wobbling bands is proposed for the first time in two even-even nuclei, $^{134}$Ce and $^{136}$Nd. The predominant $E2$ character of the $\Delta I = 1$ transitions connecting the one-phonon wobbling band in $^{134}$Ce to the two-quasiparticle yrast band supports the wobbling interpretation. The extracted wobbling frequencies decrease with increasing spin, indicating the transverse character of the wobbling motion, with the angular momenta of the two quasiparticles aligned perpendicular to the axis of collective rotation. A candidate for two-phonon wobbling motion is also proposed in $^{136}$Nd.  
The wobbling frequencies calculated in the harmonic frozen approximation are in good agreement with the experimental ones for both  the$^{134}$Ce and $^{136}$Nd nuclei.    

\end{abstract}

\begin{keyword}
\texttt{Nuclear structure of $^{134}$Ce and $^{136}$Nd}\sep model calculations\sep collective excitations \sep wobbling motion
\end{keyword}

\end{frontmatter}


\section{Introduction}

The recent studies of the Lanthanide nuclei with $A \approx 130$ were mainly driven by the search for chiral bands in odd-odd and odd-even nuclei \cite{fra}. Candidates for chiral partners were proposed in nuclei from $^{126}$Cs to $^{138}$Eu. Very recently, the transverse wobbling mode has been discovered in the odd-even $^{135}$Pr nucleus \cite{fra-trans,135pr}, in which case the unpaired $\pi h_{11/2}$ proton lies at the bottom of the $\pi h_{11/2}$ sub-shell, having therefore particle character. As a consequence, in order to maximize the overlap of the density distribution with the triaxial core, its angular momentum is along the short axis of the triaxial core, being therefore perpendicular to the angular momentum of the core which is built along the intermediate axis if one assumes hydrodynamic moments of inertia. The wobbling excitations can also occur in odd-even nuclei with an unpaired $\nu h_{11/2}$ neutron lying at the top of the $\pi h_{11/2}$ sub-shell, having therefore hole character. In this case the angular momentum of the unpaired neutron is aligned along the long axis, being again perpendicular to the collective angular momentum of the triaxial core. Such excitations have been recently identified in $^{133}$Ce \cite{133ce-wobb}. 

In the Lu-Hf region, wobbling bands were only observed in odd-even triaxial superdeformed nuclei \cite{163lu}.
Even if wobbling bands are predicted since long in even-even triaxial nuclei, no clear experimental evidence has been reported until now in configurations without unpaired nucleons. However,  the band built on the $\nu h_{11/2}^{-2}$ two-quasihole neutron configuration of the normal-deformed $^{138}$Nd was recently interpreted as an wobbling collective excitation \cite{138nd}. Even a two-phonon wobbling excitation was suggested in the same $^{138}$Nd nucleus, but the limited number of observed states and lack of angular distribution data did not allow to draw a definite conclusion.  \\
The recent evidence for wobbling motion in normal-deformed triaxial nuclei of the $A\sim130$ mass region showed that the transverse coupling is realized over a frequency range much shorter than for the triaxial superdeformed bands observed in the Lu region, but sufficient to allow experimental observation  \cite{133ce-wobb}. It is therefore natural to imagine a transverse geometry with two $h_{11/2}$ particles coupled to a triaxial core. The resulting bands would be built on the $I^{\pi}=10^+$ maximum aligned states of even-even nuclei from the $A\sim 130$ mass region. The angular momentum of two maximally-aligned $h_{11/2}$ quasiparticles in an even-even nucleus, $I=10$ $\hbar$,  being larger than that of one  $h_{11/2}$  quasiparticle in an odd-even nucleus, $I=11/2$ $\hbar$, would assure the observation of the wobbling motion over a larger frequency interval before the decoupling from the core under the effect of the Coriolis force. Rotational bands built on the $\pi h_{11/2}^2$ and $\nu h_{11/2}^{-2}$, $I=10^+$ states are known in several nuclei of the $A\sim130$ mass region \cite{138nd,134ce,136nd-plb,136nd-sun,136nd-mergel,136nd-vre}. Calculations performed using different models converge to the conclusion that these 2-qp bands are based on triaxial shapes. In addition to the bands with even spins built on the lowest $10^+$ states, also bands with odd spins which decay to the bands built on the $10^+$ states were identified in the three nuclei $^{134}$Ce, $^{136}$Nd and $^{138}$Nd \cite{138nd,134ce,136nd-plb}. These odd-spin bands were tentatively assigned as two-quasiparticle (2-qp) triaxial configurations in $^{134}$Ce and $^{136}$Nd, while in $^{138}$Nd the odd-spin band which decays to the band built on the $I^{\pi}=10^+$ isomer was interpreted as wobbling collective excitation. Excepting $^{134}$Ce, for which angular distribution results were recently published \cite{134ce}, no detailed experimental information on the angular distribution of the connecting transitions between the 2-qp odd-spin bands of $^{136}$Nd and $^{138}$Nd were reported. 

The present paper proposes a new interpretation of the 2-qp odd-spin bands of $^{134}$Ce and $^{136}$Nd in terms of transverse wobbling motion. Simple calculations using the recently introduced Harmonic Frozen Approximation (HFA) \cite{fra-trans} are used to interpret the observed wobbling frequencies. The adopted deformation of the various configurations are calculated with the Cranked Nilsson-Strutinsky (CNS) model.

\section{Experimental information}

\begin{figure}[ht]
\vskip -1.0 cm
\hskip  -. cm
\rotatebox{-0} {\scalebox{.7}{\includegraphics{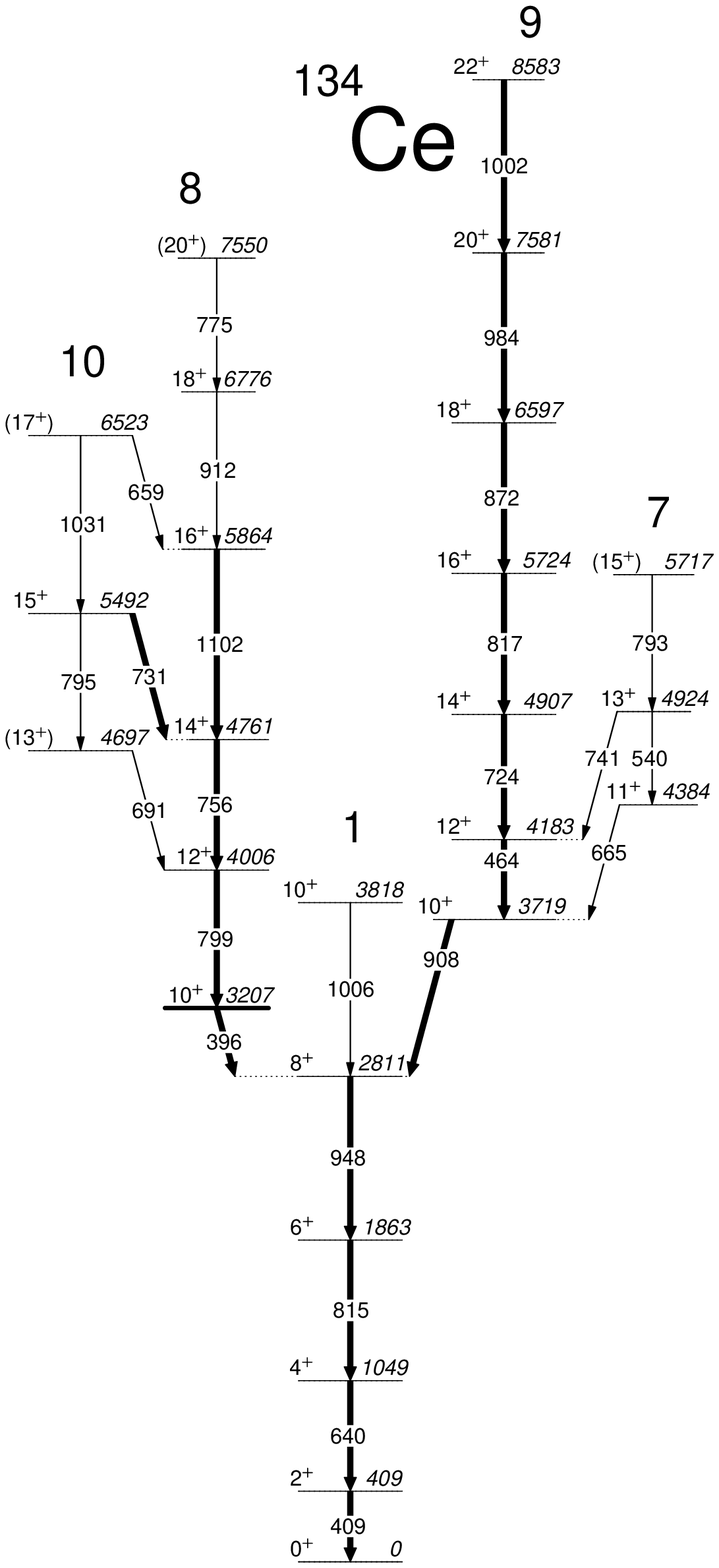}}}
\vskip -.0 cm
\caption {\label{level-scheme-1} Partial level scheme of  $^{134}$Ce showing the yrast band 1, the band 8 built on the $\nu h_{11/2}^{-2}$ configuration assigned to the $10^+$ isomer, the band 9 built on the  $\pi h_{11/2}^2$ configuration, and the odd-spin bands 7 and 10 interpreted as wobbling bands. The band labels are those used in Ref. \cite{134ce}.}
\vskip .0 cm
\end{figure}

\begin{figure}[ht]
\vskip -.0 cm
\hskip  -. cm
\rotatebox{-0} {\scalebox{.6}{\includegraphics{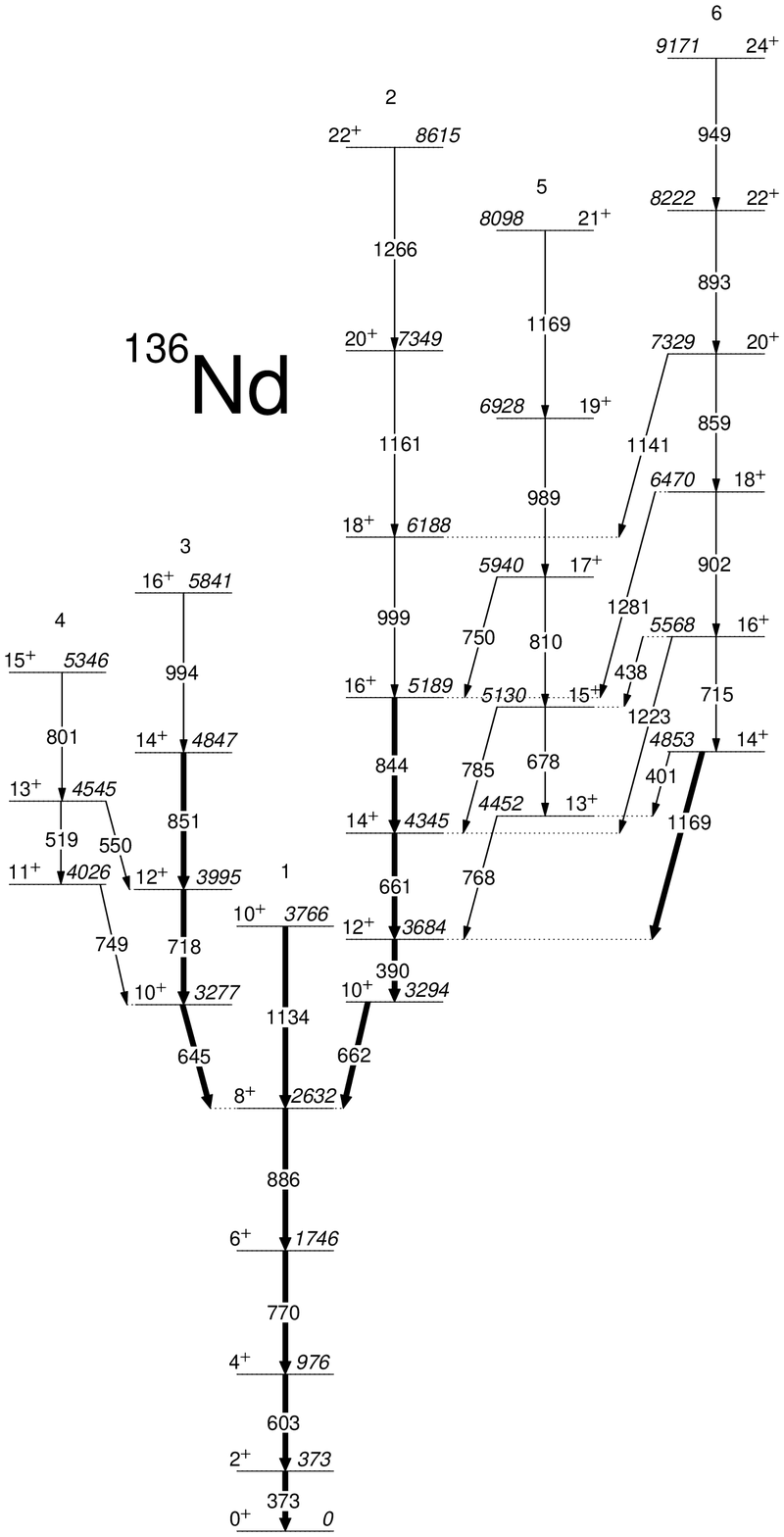}}}
\vskip -.0 cm
\caption {\label{level-scheme-2} Partial level scheme of  $^{136}$Nd showing the newly identified bands at low spin, the one- and two-phonon wobbling bands 5 and 6 which decay to the yrast 2-qp band built on the $\pi h_{11/2}^2$ configuration, and the band 4 which decays to band 3 built on the 2-qp band with $\nu h_{11/2}^{-2}$ configuration. }
\vskip .0 cm
\end{figure}

The wobbling interpretation is based on the behavior of the wobbling frequency, on the recently published mixing ratios of the  $\Delta I=1$ connecting transitions between the bands 7 and 9 of $^{134}$Ce \cite{134ce}, and on the similarities between the observed 2-qp bands of $^{134}$Ce and $^{136}$Nd. The partial level schemes showing the discussed bands of $^{134}$Ce and $^{136}$Nd are given in Figs. \ref{level-scheme-1} and \ref{level-scheme-2}.

The bands 7 and 10 of $^{134}$Ce have been recently reported in Ref. \cite{134ce}. The absolute values of the mixing ratios extracted from the angular distribution of the 664- and 741-keV, $\Delta I=1$ transitions are significantly larger than one, 2.13(48) and 1.73(52) respectively, clearly indicating their predominant $E2$ character, with 82(5)\% and 75(6)\% $E2$ component, respectively. We identified two new weak transitions of 691 and 795 keV populating and de-exciting the $(13^+)$ state of band 10, respectively. 

The bands 5 and 6 of $^{136}$Nd have been reported in Ref. \cite{136nd-plb,136nd-sun}, while the bands 3 and 4  have been reported in Refs. \cite{136nd-sun,136nd-mergel}. Some of the transitions reported in the previous papers are drawn differently in Fig. \ref{level-scheme-2}, to reveal the band pattern. We identified two new weak transitions of 401 and 438 keV de-exciting the $14^+$ and $16^+$ states of band 6, respectively. The experimental mixing ratios of the decaying transitions of bands 4, 5 and 6 could not be extracted due to limited statistics.

\section{Discussion}

\begin{figure}[]
\vskip -.0 cm
\hskip  -. cm
\rotatebox{-90} {\scalebox{.25}{\includegraphics{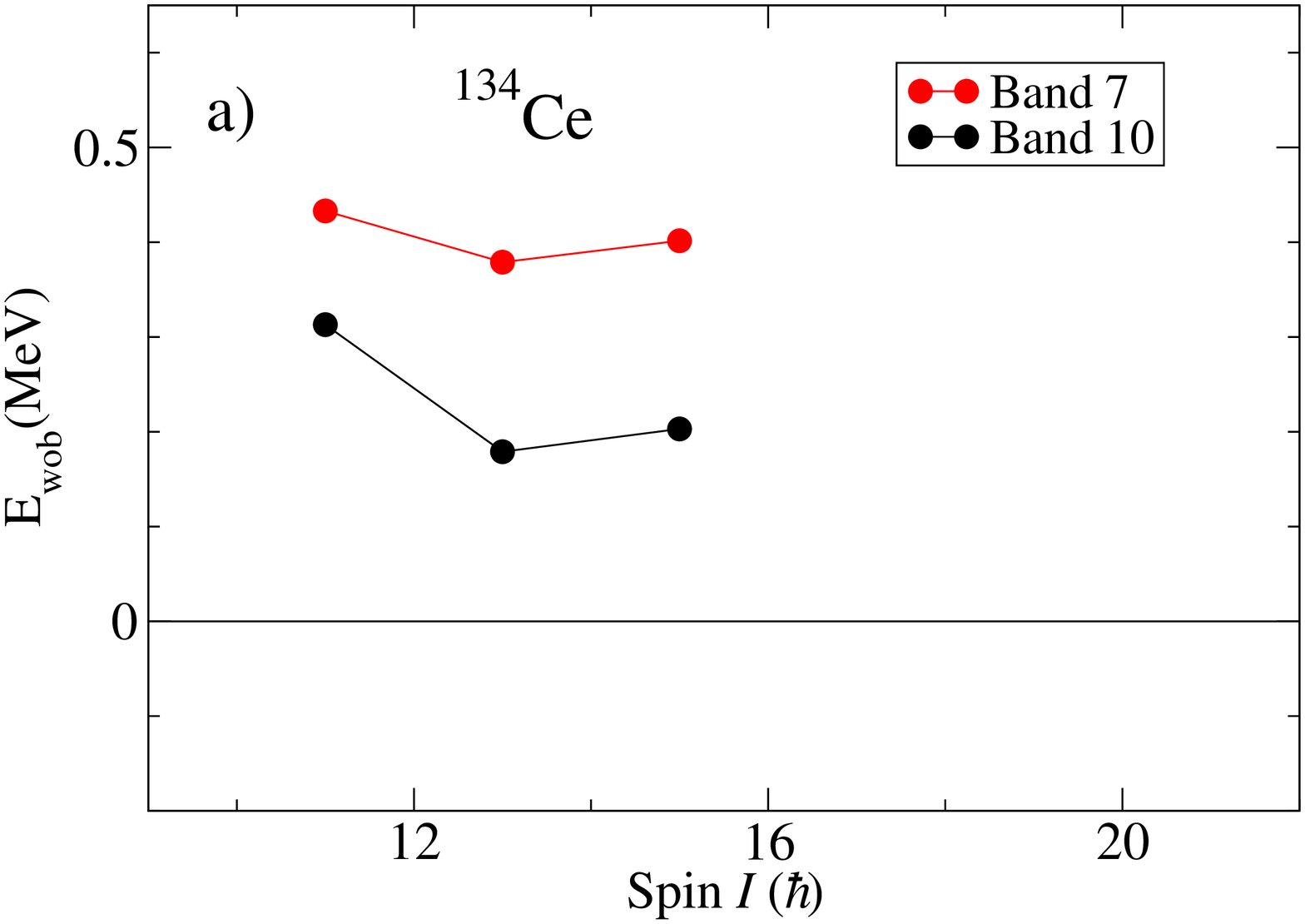}}}
\rotatebox{-90} {\scalebox{.25}{\includegraphics{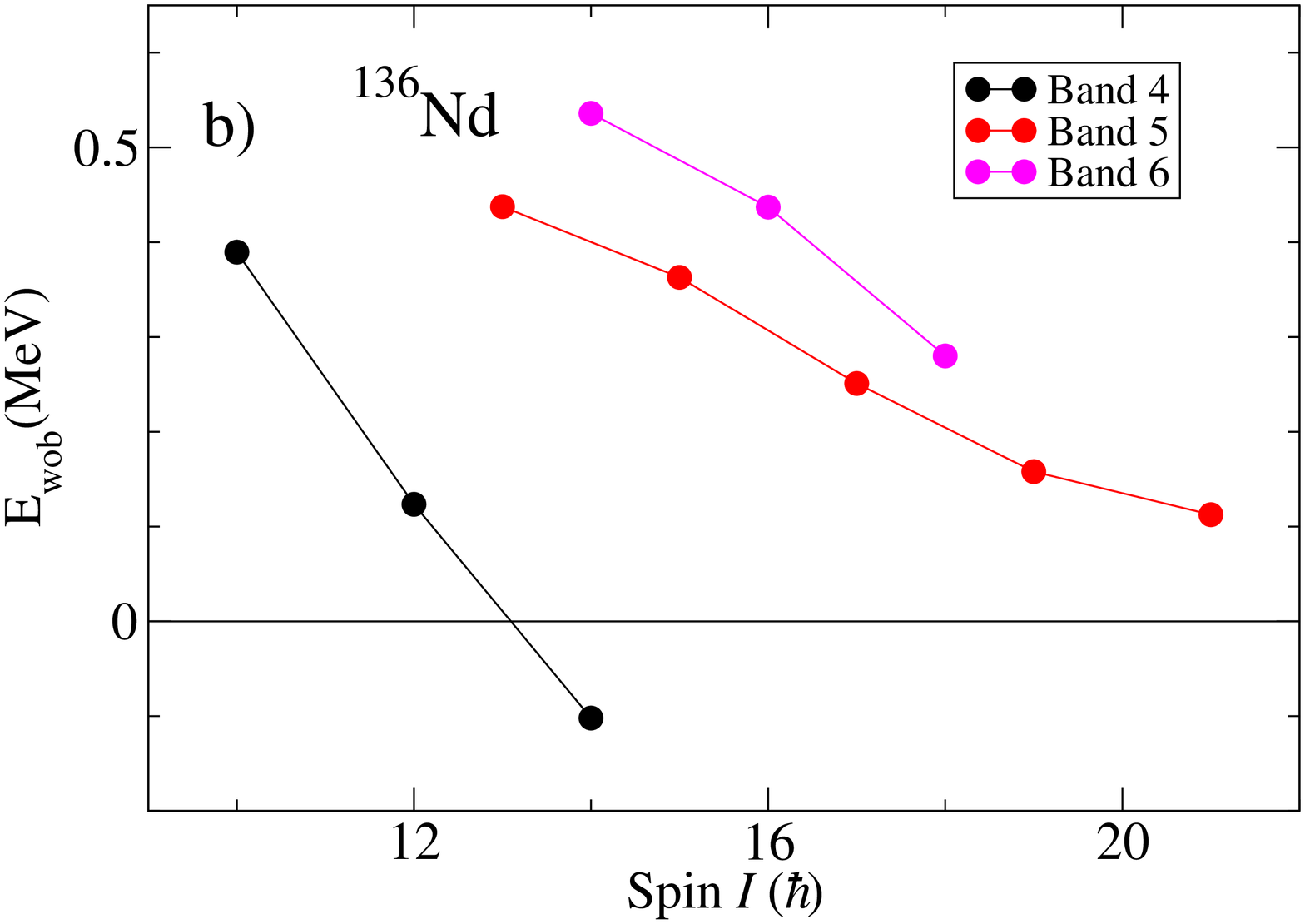}}}
\vskip -.0 cm
\caption{(Color online) Wobbling phonon energy (wobbling frequencies) of the candidates for one- and two-phonon wobbling bands in a) $^{134}$Ce and b) $^{136}$Nd. }
\label{e-wobb-exp} 
\vskip .0 cm
\end{figure}

The wobbling energies (also called wobbling frequencies) of the 1-phonon  ($n_{w}=1$) band defined as the energy difference with respect to the interpolated energies of the $n_{w}=0$ band 
\begin{equation}
E_{wobb}(I)=E(I,n_w=1)- 
[E(I+1,n_w=0)+E(I-1,n_w=0)]/2
\end{equation}
are shown in Fig. \ref{e-wobb-exp}. 
One can observe the decrease of the wobbling frequencies with increasing spin, which is characteristic for the transverse wobbling mode proposed theoretically in Ref. \cite{fra-trans} and observed experimentally in $^{135}$Pr \cite{135pr}. This behavior, together with the predominant $E2$ character of the $\Delta I=1$ decaying transitions of band 7 in $^{134}$Ce strongly support the transverse wobbling interpretation. No detailed angular distribution and polarization information is available for the other bands of $^{134}$Ce and $^{136}$Nd discussed in the present paper, and therefore the electromagnetic character of the transitions is uncertain. The wobbling interpretation of these bands has to wait a definite confirmation from future  experiments dedicated to the measurement of the electromagnetic character of the $\Delta I=1$ connecting transitions.  

The extracted wobbling frequencies for the bands reflect their different configurations. The two bands of $^{134}$Ce based on the 3207-keV and 3719-keV, $I=10^+$ states are assigned to the same $\nu h_{11/2}^{-2}$ configuration, but corresponding to coexisting shapes with different deformations \cite{134ce-zemel}, or to low-$K$ configurations built on the ground-state band and the $\gamma$-band \cite{sheikh}. The yrast $\nu h_{11/2}^{-2}$ configuration calculated by the CNS model can be naturally assigned to band 10 based on the yrast $I=10^+$ state. Its deformation changes from $(\varepsilon_2 \sim 0.14, \gamma \sim -35^{\circ})$ at spin $I=12$, to $(\varepsilon_2 \sim 0.12, \gamma \sim -49^{\circ})$ at spin $I=22$ \cite{134ce}. The deformation of the non-yrast $\nu h_{11/2}^{-2}$ configuration that can be assigned to band 7 built on the 3719-keV,  $I=10^+$ state cannot be calculated within the CNS model. Based on  the similarity with bands observed in many neighboring nuclei, in particular with band 2 of the presently studied $^{136}$Nd nucleus, one can assume that the deformation of band 7 is similar to that of the $\pi h_{11/2}^{2}$ configuration assigned to band 2 in $^{136}$Nd, which changes from $(\varepsilon_2 \sim 0.21, \gamma \sim 25^{\circ})$ at spin $I=12$, to $(\varepsilon_2 \sim 0.19, \gamma \sim 25^{\circ})$ at spin $I=22$. One can therefore make a step further in the interpretation of the bands based on the two lowest $I=10^+$ states in $^{134}$Ce, combining the two previously interpretations in terms of coexisting shapes with different deformations \cite{134ce-zemel} and bands built on the ground-state band and the $\gamma$-band \cite{sheikh}: the band based on the yrast $I=10^+$ state is based on a configuration with lower deformation and the unpaired neutrons aligned along the long axis ($\gamma < 0^{\circ}$), while the band based on the non-yrast $I=10^+$ state is based on a configuration with higher deformation and the unpaired neutrons aligned along the short axis ($\gamma > 0^{\circ}$).

The configurations assigned to the bands 7 and 10 of $^{134}$Ce involve therefore two neutron holes in orbitals from the top of the $h_{11/2}$ sub-shell, which, in order to maximize the overlap of the density distribution with the triaxial core, are aligned parallel to either the short or the long axes, respectively. As the hydrodynamical moments of inertia (MoI) are largest for rotation along the intermediate axis, the angular momenta of the two quasiholes are in both cases perpendicular to the angular momentum of the core, corresponding to transverse wobbling. As this is the first time that transverse wobbling bands are proposed in even-even nuclei, we will call them  {\it 2-qp transverse wobbling bands}, to be distinguished from the {\it 1-qp transverse wobbling bands} observed in odd-even nuclei \cite{135pr,133ce-wobb}. However, one note that the wobbling mode built on 2-qp configurations has been recently proposed for a similar $\nu h_{11/2}^{-2}$ configuration in $^{138}$Nd, but without invoking the transverse coupling geometry \cite{138nd}.

Based on the similarity between the bands 10 of $^{134}$Ce and 4 of $^{136}$Nd, one can also interpret band 4 of $^{136}$Nd as a  {\it 2-qp transverse wobbling band} built on the $\nu h_{11/2}^{-2}$ configuration assigned to band 3 towards which it decays  \cite{136nd-paul}. As one can see in Fig. \ref{e-wobb-exp}, the wobbling phonon energies of bands 10 of $^{134}$Ce and 4 of $^{136}$Nd are smaller than those of band 7 of $^{134}$Ce, and 5 and 6 of $^{136}$Nd, and decrease more sharply with spin, like in the case of the 1-qp wobbling bands in $^{133}$Ce in which one $h_{11/2}$ neutron hole is coupled in a transverse geometry with the triaxial core \cite{133ce-wobb}. The decrease of the wobbling frequency in band 10 of $^{134}$Ce is observed for the two lowest spins, being followed by an increase which signals the decoupling from the transverse geometry.  

Band 5 of $^{136}$Nd decays to band 2 which has a $\pi h_{11/2}^2$ configuration \cite{136nd-paul}. The wobbling frequency is larger and decreases more slowly than in band 4. The calculated deformation within the CNS formalism changes from $(\varepsilon_2 \sim 0.21, \gamma \sim 25^{\circ})$ at spin $I=12$, to $(\varepsilon_2 \sim 0.19, \gamma \sim 25^{\circ})$ at spin $I=22$, which is therefore larger than that of the $\nu h_{11/2}^{-2}$ configuration assigned to band 10 of $^{134}$Ce and band 4 of $^{136}$Nd. 

Band 6 of $^{136}$Nd decays to both band 2 and band 5. Its wobbling frequency is higher than that of band 5 and has the same gradual decrease as that observed in band 5. It is a good candidate for a two-phonon 2-qp wobbling band. 

\begin{figure}[]
\vskip -.0 cm
\hskip  -. cm
\rotatebox{-90} {\scalebox{.25}{\includegraphics{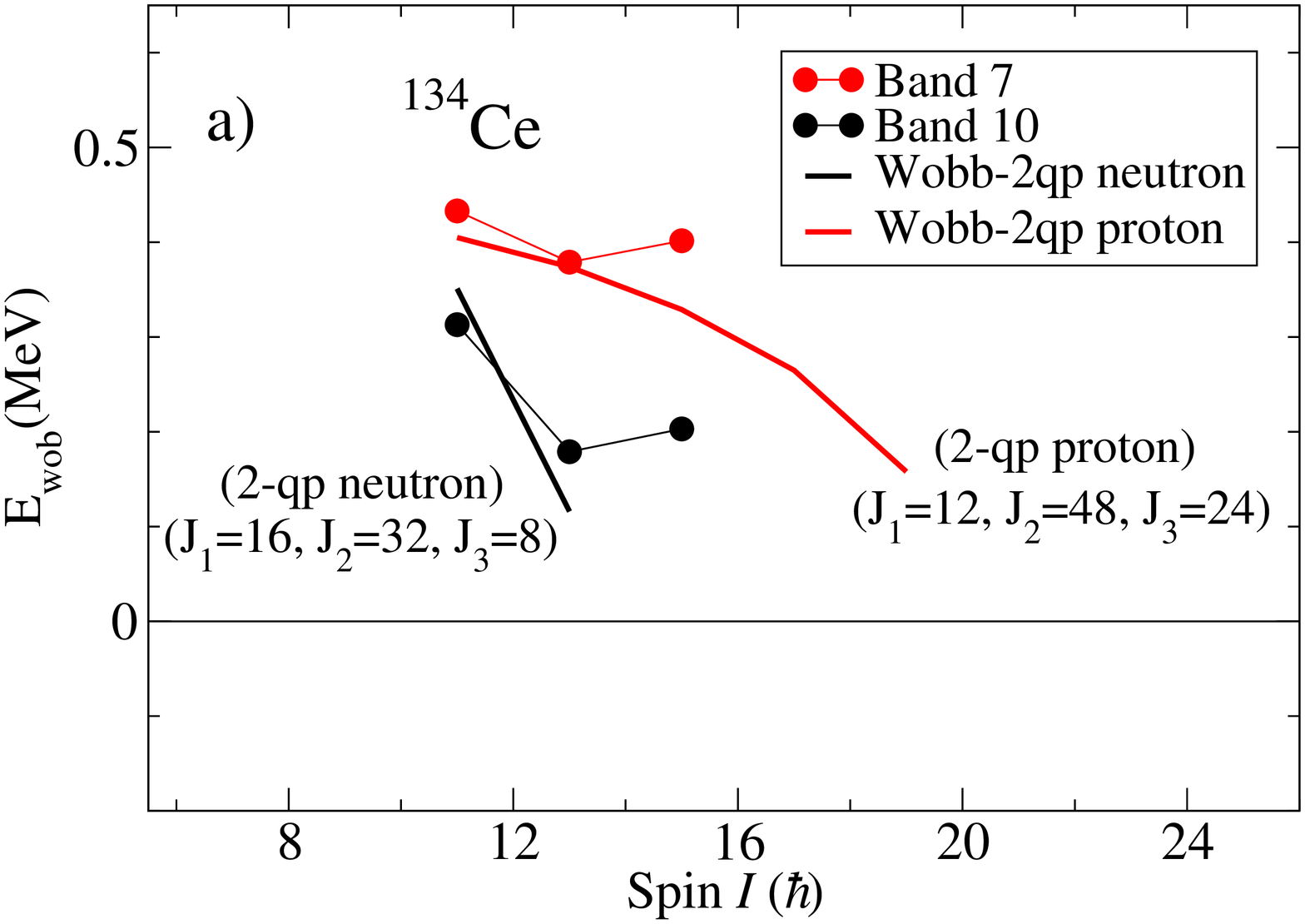}}}
\rotatebox{-90} {\scalebox{.25}{\includegraphics{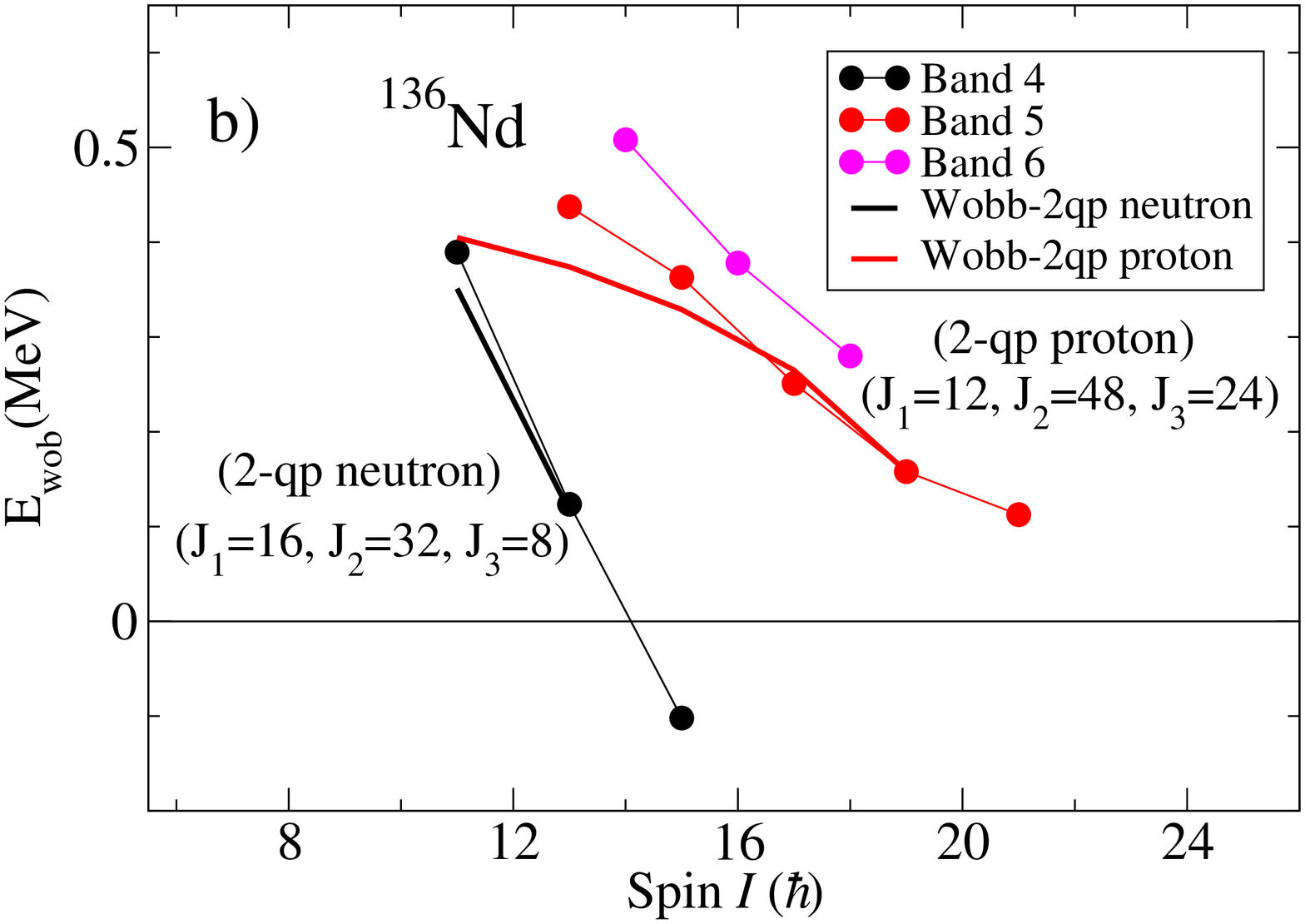}}}
\vskip -.0 cm
\caption{(Color online) Comparison between the experimental and calculated HFA wobbling frequencies of the candidate wobbling bands in a) $^{134}$Ce and b) $^{136}$Nd. The calculated wobbling frequencies are drawn with continuous thick line. }
\label{e-wobb} 
\vskip .0 cm
\end{figure}

To further investigate if the 2-qp transverse wobbling interpretation of the bands observed in $^{134}$Ce and $^{136}$Nd is realistic, we calculated the wobbling frequencies using the Harmonic Frozen Approximation, in which the angular momenta of the unpaired nucleons are frozen perpendicular to the collective angular momentum of the core \cite{fra-trans}. The wobbling frequency depends on the three MoI's and on the spins of the unpaired quasiparticles. It is reasonable to assume that the two unpaired $h_{11/2}$ quasiparticles are coupled to maximum angular momentum $I=10$. For the MoI's of the $\nu h_{11/2}^{-2}$ configuration we can adopt values similar to those calculated using the RPA model applied to 2-qp configurations employed for $^{138}$Nd \cite{138nd}: $\mathcal{J}_1=16$, $\mathcal{J}_2=32$ and $\mathcal{J}_3=8$ (in $\hbar^2$/MeV). Note that  $\mathcal{J}_1$, $\mathcal{J}_2$ and $\mathcal{J}_3$ correspond to rotations around the long, intermediate and short axes, or alternatively around the 1-axis, 2-axis and 3-axis, respectively. For the MoI's of the $\pi h_{11/2}^{2}$ configuration we adopt larger values, which correspond to the overall larger deformation calculated within the CNS model : $\mathcal{J}_1=12$, $\mathcal{J}_2=48$ and $\mathcal{J}_3=24$ (in $\hbar^2$/MeV). In this case $\mathcal{J}_1 < \mathcal{J}_3$, which is opposite to the relation  $\mathcal{J}_1 > \mathcal{J}_3$ valid for the $\nu h_{11/2}^{-2}$ configuration. These moments of inertia have to be considered as effective moments of inertia, which are larger than the collective ones by the contribution of the unpaired quasiparticles, which are two neutron holes for the 1-axis and two proton particles for the 3-axis. The ratio of the MoI's for the axes without aligned particles is obtained from the hydrodynamical model with maximal triaxiality $\gamma=30^{\circ}$: $\mathcal{J}_2/\mathcal{J}_3=32/8=4$ for the $\nu h_{11/2}^{-2}$ configuration or $\mathcal{J}_2/\mathcal{J}_1=48/12=4$ for the $\pi h_{11/2}^{2}$ configuration. The contribution  of the unpaired quasiparticles to the effective MoI is taken as large as the hydrodynamical MoI corresponding to the axis with aligned quasiparticles: 8 $\hbar^2$/MeV for the  $\nu h_{11/2}^{-2}$ configuration and 12 $\hbar^2$/MeV for the  $\pi h_{11/2}^{2}$ configuration. As one can see in Fig. \ref{e-wobb}, an overall good agreement with the experimental values is obtained. Of course, the present schematic calculations have to be confirmed in the future by more realistic and detailed calculations like RPA built on 2-qp configurations.

 
\section{Summary}

One-phonon and possible two-phonon transverse wobbling bands are proposed for the first time in the even-even nuclei $^{134}$Ce and $^{136}$Nd. The experimental wobbling frequencies decrease with increasing spin, indicating the transverse character of the wobbling motion, with the angular momenta of the two quasiparticles perpendicular to the collective angular momentum of the triaxial core. A candidate for two-phonon wobbling motion is also proposed in $^{136}$Nd.  
The calculated wobbling frequencies in the harmonic frozen approximation are in good agreement with the experimental values for both $^{134}$Ce and $^{136}$Nd nuclei. 
To give a solid support to the present wobbling interpretation, future experiments to determine the electromagnetic character of the $\Delta I=1$ connecting transitions and the reduced transitions probabilities are necessary. Detailed calculations like RPA built on 2-qp configurations are also needed to confirm the present schematic calculations.

\bibliography{mybibfile}

\end{document}